\documentclass [10pt,twoside]{article}

\usepackage{shrthnds}
\usepackage{cite}
\usepackage{graphicx}
\usepackage[usenames]{color}
\usepackage{subfigure}
\usepackage{setspace}

\usepackage{multirow}

\usepackage[hang,small]{caption}

\usepackage{amsmath}                  %text inside eqnarray

\usepackage{geometry}
\usepackage{fancyhdr}
\usepackage{titlesec,titletoc}
  \titleformat{\section}{\Large\sf\bfseries}{\thesection}{1em}{}
  \titleformat{\subsection}{\large\sf\bfseries}{\thesubsection}{1em}{}

\title{\sf\bfseries \ntitle}
\author{\normalsize  Naveen K. Singh\footnote{email: naveenks@prl.res.in}}
\date{}%{\today}

\newcommand{\pghdr}{\footnotesize {Naveen K. Singh}  -- Unimodular Constraint on global scale invariance}
\newcommand{\ntitle}{Unimodular Constraint  on   global scale Invariance }

\begin{document}
\vspace{-3cm}
\maketitle
%\vspace{-1.5cm}
\vspace{-0.6cm}
\bc
\small{ Theory Division, Physical Research Laboratory, Navrangpura, Ahmedabad - 380 009, India}
%{\small 1) Aff 1\\
%2) Aff 2\\
%3) Aff 3}
\ec

\bc\begin{minipage}{0.9\textwidth}\begin{spacing}{1}{\small {\bf Abstract:}
%%%%%%%%%%%%%%%%%%%%%%%%%%%%%%%%%%%%%%%%%%%%%%%%%%%%%%
% ABSTRACT
%%%%%%%%%%%%%%%%%%%%%%%%%%%%%%%%%%%%%%%%%%%%%%%%%%%%%%
We study a model with global scale invariance within the framework
of unimodular gravity. The global
scale invariant gravitational action which follows the unimodular
general coordinate transformations is considered without invoking any scalar field. This is generalization of conformal theory
described in the Ref. \cite{Mannheim}. The possible
solutions for the gravitational potential under static linear field approximation are discussed.
The new modified solution has additional corrections to the Schwarzschild solution which describe the galactic
rotational curve. A comparative study of unimodular theory with conformal theory is also presented. Furthermore, the
cosmological solution is studied and it is shown that the unimodular constraint preserve the de
Sitter solution explaining the dark energy of the universe.
 
% estimate the integration constants for the best fit to the galactic rotational curve.

%%%%%%%%%%%%%%%%%%%%%%%%%%%%%%%%%%%%%%%%%%%%%%%%%%%%%%
}\end{spacing}\end{minipage}\ec

\section{Introduction}
Surprisingly, we live in the era of late time acceleration of universe \cite{Perlmutter,Riess}. The acceleration in expansion of universe offers
the possibility of new physics of the unknown component ``dark energy''. The dark matter, whose physical nature is only partially known, is also
one of the interesting subjects of research in cosmology. Dark matter was introduced by
Fritz Zwicky in year 1933 \cite{Turner,Einasto} to account for evidence of “missing mass” in Coma cluster. The
more interesting issue to notice is that dark matter and dark energy dominate the energy
density of the universe. The observed dark energy and dark matter contribute approximately
 $72.8 \%$ and $22.7\%$ respectively to the total energy content of the universe. There are number of 
scalar field theories such as Quintessence, K-essence, Chaplygin gas model  and 
theory of modified gravity such as  $f(R)$, DGP model,
 etc. to describe the dark side of universe. In the recent literature \cite{Chile,GRL,CCBP}, authors discuss the absence of the dark matter
in the vicinity of solar system. The  dark matter might be internal property of the space and hence might be explained by 
the modified theory of gravity\cite{Mannheim,Capo1,Iorio_Rugg,Martins_Salucci
,Saffari_Sob,Zhao_Li,Soti_Faraoni}. In this paper, such a
 theory of modified gravity is considered which has global scale invariance along with the unimodular constraint to explain
the galactic rotational curve and acceleration of universe expansion. The scale symmetry 
prevents any dimensionful parameter in the action and
hence might give a possible explanation for the cosmological constant problem
\cite{Mannheim1,Mannheim2,Deser,Shaposhnikov:2008a,Shaposhnikov:2008b,JMS,JM10_2}. This symmetry is broken if the theory contains any dimensionful parameters,
such as particle masses, cosmological constant, gravitational constant etc..  Although scale invariance is generally believed to be anomalous, it is
possible to maintain this symmetry in the full quantum theory if the symmetry is broken
by a soft mechanism \cite{Shaposhnikov:2008a,Shaposhnikov:2008b,JMS,JM10_2}. 
It has been shown  \cite{JMS,AJS,ChengPRL,JM10_2} that scale
 invariance has its other  advantages in explaining the dark sector of the universe. The local scale invariance explains the current era of 
the universe \cite{AJS}, favoring the $\Lambda CDM$ model. In  \cite{AJS,ChengPRL} the symmetry breaking mechanism 
of local scale invariance generates the dark energy and vector field which  acts 
as dark matter field. The global  scale transformation is given by 
\ba
x^\mu &\rar & \Lambda x^{\mu} \nonumber 
% ds^2 &\rar & \Lambda^2 ds^2
% \bar g_{\mu\nu} &\rar & \bar g_{\mu\nu}\nonumber\cr
\ea
where  $\Lambda$ is a constant parameter. 
 The above transformations make the action like $\int R^2 \sqrt{-g}d^4 x$ or $\int \phi^2 R \sqrt{-g} d^4 x$ invariant.\\
The theories of higher order invariants in the action were initially started by Weyl in 1919 and by Eddington in 1923.  One  of
the advantage of $f(R)$ theory is it describes the early
universe \cite{Starobinsky,Starobinsky:1983zz,Hwang_Noh}. Further, it also explains the late time acceleration \cite{Soti_Faraoni,Capo_Car_Tro} in the 
expansion of universe and might be an alternative for the dark matter
\cite{Soti_Faraoni,Mannheim,Capo1,Iorio_Rugg,Martins_Salucci,Saffari_Sob,Zhao_Li}. The theory follows 
the principle of covariance similar as the Einstein-Hilbert action. 
It modifies the Einstein equation as well as  the gravitational
potential.  The only quadratic 
terms of the curvature scalar in the action  preserves the global scale invariance.  It is reasonable to take only quadratic
terms of the curvature scalar in the action, since the solution could include the Schwarzschild solution and the corrections 
in addition \cite{Mannheim}.  Using the Gauss-Bonnet identity, we may write such global scale invariant action as \cite{Donoghue}
\ba
-\alpha_g \int (R^{\mu\nu}R_{\mu\nu} + \gamma R^2) \sqrt{-g}d^4 x \label{Quad_action}
\ea
In this paper we generalize the conformal theory \cite{Mannheim} by imposing the unimodular
constraint. Conformal theory is one of the 
special case of (\ref{Quad_action}), where $\gamma=-1/3$  \cite{Mannheim}.  In  
subsection (\ref{Field_eqn}), the field equation 
of metric is derived. In  section {\ref{Field_eqn_uni}},  the 
corresponding equation for the unimodular theory is given. The field equation
is solved under static linear field approximation and  corresponding galactic rotational curves are discussed in 
section  \ref{Field_eqn_uni}.  The cosmological solution with unimodular constraint is discussed 
in  section (\ref{Cosmo_sol}). The last section (\ref{conclusion}) contains the discussion and  conclusions.
\subsection{Field Equation} \label{Field_eqn}
The variation of action \ref{Quad_action} gives the field equation as
% \ba
% \delta S &=& -\alpha_g \int \bigg[ \sqrt{-g} (2 \delta g^{\rho \mu} g^{\sigma \nu} R_{\rho\sigma} R_{\mu\nu} + 2 g^{\rho \mu} g^{\sigma \nu}
% R_{\rho\sigma} \delta R_{\mu\nu} +2 \gamma R(\delta g^{\mu\nu} R_{\mu\nu}+ g^{\mu\nu}\delta R_{\mu\nu})) \nonumber \\
% &-&\frac{1}{2}\sqrt{-g}g_{\mu\nu}\left(\gamma R^2 + R_{\rho\sigma}R^{\rho\sigma}\right)\delta g^{\mu\nu}\bigg]d^4 x 
% \ea
% Now using the identity \cite{Landua_Lif},
% \ba
%  \delta R_{\mu\nu} = \bigg[
% (\delta\Gamma^{\lambda}_{\mu\nu})_{; \lambda} - (\delta\Gamma^{\lambda}_{\mu\lambda})_{; \nu}\bigg],
% \ea 
\ba
W_{\mu\nu}^{1} + W_{\mu\nu}^{2}=0 \ .
\ea
Here $W_{\mu\nu}^{1}$ and $W_{\mu\nu}^{2}$ are the terms corresponding to the variation
of $R^2$ and $R_{\rho\sigma} R^{\rho\sigma}$  and these are given as
\ba
W_{\mu\nu}^{1} &=& -\frac{\gamma}{2} R^2 g_{\mu\nu}  + 2 \gamma \bigg[g^{\alpha\beta} R_{; \alpha ;\beta} g_{\mu\nu}- R_{;\mu;\nu}\bigg] 
+ 2 \gamma R R_{\mu\nu} ,\nonumber \\
W_{\mu\nu}^{2}&=& -\frac{1}{2} R_{\rho\sigma}R^{\rho\sigma} g_{\mu\nu} + 2 R_{\mu\rho}R_{\nu\sigma}g^{\rho\sigma} 
-2 (R^{\alpha\beta})_{;\mu;\beta} g_{\alpha\nu} + (R_{\mu\nu})_{;\rho}^{;\rho} + (R^{\alpha\beta})_{;\beta;\alpha}g_{\mu\nu} 
\ea 
respectively.
\section{Unimodular Gravity} \label{Field_eqn_uni}
Unimodular gravity was introduced in  \cite{Anderson,Einstein} and has been
reviewd in \cite{Weinberg}. The theory is subclass of general theory of
relativity but with a constraint in addition, i.e., the 
determinant of the metric is not dynamical; $g_{\mu\nu}\delta g^{\mu\nu} =0$. The motivation of the unimodular 
gravity is to solve the cosmological constant problem as we don't have any such term in the action.  Further, in the reference \cite{Jain:2011},
authors discuss dynamics of expansion of universe with the unimodular theory of gravity taking the dynamical part of determinant of metric
as a separate scalar field. However, in this paper, any scalar field is not considered. The condition $g_{\mu\nu}\delta g^{\mu\nu} =0$ modifies the Einstein equation as following \cite{Anderson,Weinberg},
\ba
R_{\mu\nu}-\frac{1}{4}g_{\mu\nu} R =  \kappa \left( T_{\mu\nu} -\frac{1}{4} g_{\mu\nu}T\right) \label{Einstein_uni} \ ,
\ea
where $\kappa$ is coupling constant, $T_{\mu\nu}$ is energy momentum tensor of source field and $T$ is its trace. The 
Eq. \ref{Einstein_uni} is traceless part of the Einstein equation. The variation of action \ref{Quad_action} gives following field
equation
\ba
W_{\alpha\beta} &=& -\frac{1}{2}g_{\alpha\beta}\left(\gamma R^2 + R_{\rho\sigma}R^{\rho\sigma}\right) + 2 \gamma \bigg[g^{\mu\nu}(R)_{;\mu;\nu}g_{\alpha\beta}-
(R)_{;\alpha; \beta}\bigg] + 2 R_{\alpha\nu}R_{\beta\rho}g^{\rho\nu} \nonumber \\
&-&2 \left(R^{\mu\nu}\right)_{;\alpha;\nu}g_{\mu\beta} +
\left(R^{\mu\nu}\right)_{;\lambda}^{;\lambda} g_{\mu\alpha}g_{\nu\beta}+ \left(R^{\mu\nu}\right)_{;\nu;\mu} g_{\alpha\beta}
+ 2 \gamma R R_{\alpha\beta} =0 \label{field_eq1} \ .
\ea
The same procedure of constraint of unimodular gravity over the action given in Eq. \ref{Quad_action} gives the following field equation,
\ba
W_{\alpha\beta}^{uni}&=& -\frac{1}{2}g_{\alpha\beta}\left(\gamma R^2 + R_{\rho\sigma}R^{\rho\sigma}\right) + 2 \gamma \bigg[g^{\mu\nu}(R)_{;\mu;\nu}g_{\alpha\beta}-
(R)_{;\alpha; \beta}\bigg] + 2 R_{\alpha\nu}R_{\beta\rho}g^{\rho\nu} \nonumber \\
&-&2 \left(R^{\mu\nu}\right)_{;\alpha;\nu}g_{\mu\beta} +
\left(R^{\mu\nu}\right)_{;\lambda}^{;\lambda} g_{\mu\alpha}g_{\nu\beta}+ \left(R^{\mu\nu}\right)_{;\nu;\mu} g_{\alpha\beta}
+ 2 \gamma R R_{\alpha\beta} - \frac{W}{4} g_{\alpha\beta} =0 \label{field_eq2} \ ,
\ea 
where $W=W^\alpha_\alpha$ is trace of tensor $W_{\alpha\beta}$. The Eq. \ref{field_eq2} is the traceless part of Eq. \ref{field_eq1}. Here
 $g_{\mu\nu}\delta g^{\mu\nu} = 0$ is used, i.e., the action does not have any constant term.
\subsection{ Vacuum Solution for the Conformal Theory}
For $\gamma=-1/3$, $1/B$ of $W^{r r}$ component of Eq. (\ref{field_eq2}) gives the following equation
\ba
\frac{B'B'''}{6} -\frac{B''^{2}}{12}-\frac{B B'''}{3r}
+\frac{B'B''}{3r}-\frac{B B''}{3r^2} - \frac{B'^2}{3r^2} \nonumber \\
+\frac{2 B B'}{3 r^3}-\frac{B^2}{3 r^4}+\frac{1}{3 r^4} =0 \label{conformal} \ ,
\ea
where the metric is given by,
\ba
ds^2=- B(r) dt^2 + \frac{1}{B(r)}dr^2 + r^2 d\Omega  \label{metric} \ .
\ea
The exact vacuum of Eq. (\ref{conformal}) may be written as \cite{Mannheim}
\ba
B(r)= 1-\frac{C_1 (2-3 C_1 C_2)}{r} -3 C_1 C_2 + C_2 r -C_3 r^2 \label{conf_sol}\ ,
\ea
where, $C_1$, $C_2$ and $C_3$ are constants. Now in next subsection we generalize this for general $\gamma$.
\subsection{Vacuum Solution for Unimodular Gravity}
\begin{figure}[!t]
\begin{center}
\includegraphics[width=1.00\textwidth]{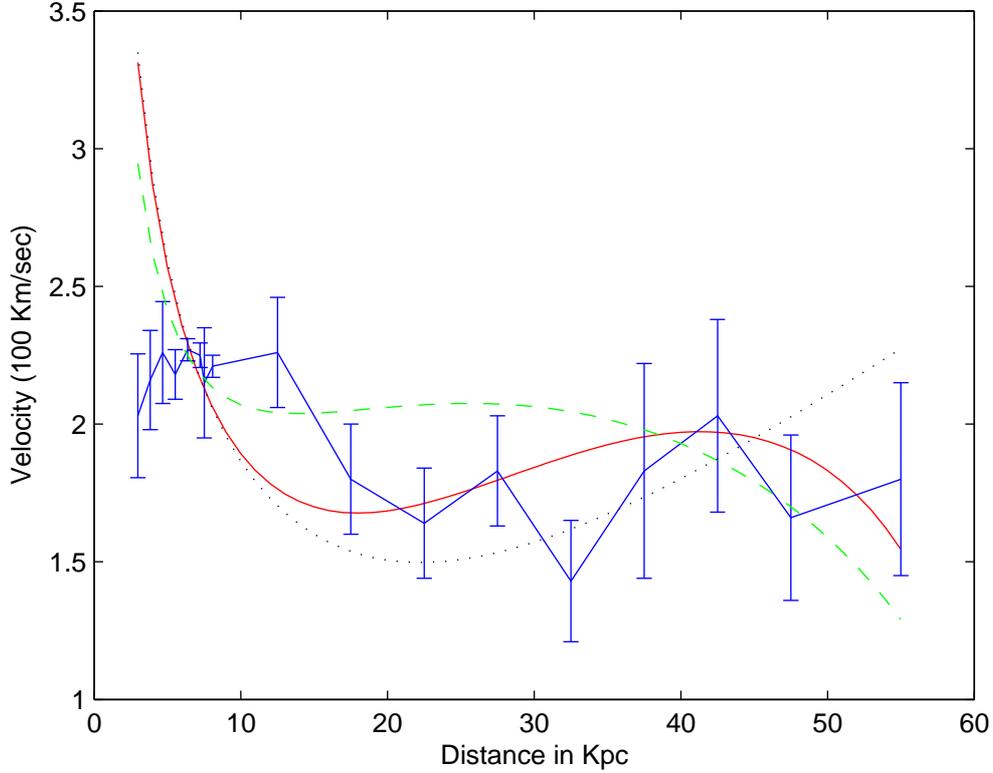}
\end{center}
\caption{The variation of velocity with distance $r$.  The dotted curve is plot of solution given by (\ref{sol2}) or (\ref{sol3}) whereas
 solid curve is for the solution (\ref{sol1}). The dashed curve is for the solution (\ref{conf_sol}). Data for Milky Way
 Galaxy is shown with the error bar. }
\label{fig:vl}
\end{figure}

In this subsection, we solve for the gravitational potential with unimodular gravity considering the  line element (\ref{metric}).
$-(1/B)$ of $t-t$ component, $-B$ of $r-r$ component and $1/r^2$ of $\theta-\theta$ component of field Eq. \ref{field_eq2} are given by 
\ba
(1+2\gamma)\frac{B''^2}{4} -\gamma \frac{B'^2}{r^2} +(1+4\gamma)\frac{B' B''}{r} -\frac{(1+2\gamma)}{r^4}
- (1+4b)\frac{B^2}{r^4} \nonumber \\ 
+ (2+6\gamma)\frac{B}{r^4}-(2+4\gamma)\frac{B B'}{r^3} +  \gamma \frac{B' B'''}{2}-(2+3\gamma)\frac{B B'''}{r} \nonumber \\
 -\gamma \frac{B B''}{r^2} -(1+\gamma)\frac{B B''''}{2} + 2(1+3\gamma)
\frac{B'}{r^3} =0 \label{ttcompt} \ ,
\ea
\ba
-(1+2\gamma)\frac{B''^2}{4} + \gamma \frac{B'^2}{r^2} -(1+4\gamma)\frac{B' B''}{r} +\frac{(1+2\gamma)}{r^4}
- (7+20 \gamma)\frac{B^2}{r^4}\nonumber \\
 + (6+18 \gamma)\frac{B}{r^4} + (2+4\gamma)\frac{B B'}{r^3} -  \gamma \frac{B' B'''}{2}-(2+5\gamma)\frac{B B'''}{r} \nonumber \\
  + (4+13\gamma) \frac{B B''}{r^2} -(1+3 \gamma)\frac{B B''''}{2} - 2(1+3\gamma)
\frac{B'}{r^3} =0 \label{rrcompt}
\ea
and 
\ba
-(1+2\gamma)\frac{B''^2}{4} + \gamma \frac{B'^2}{r^2} -(1+4\gamma)\frac{B' B''}{r} +\frac{(1+2\gamma)}{r^4}
- (3+8 \gamma)\frac{B^2}{r^4}\nonumber \\
 + (2+6 \gamma)\frac{B}{r^4} + (2+4\gamma)\frac{B B'}{r^3} -  \gamma \frac{B' B'''}{2}-\gamma \frac{B B'''}{r} \nonumber \\
  + (2+7\gamma) \frac{B B''}{r^2} -\gamma\frac{B B''''}{2} - (2+6\gamma)\frac{B'}{r^3} =0 \label{thetacompt}
\ea
respectively. Now, considering linear approximation, i.e., $ B(r) \approx 1+ \phi(r)$, we have following three equations
\ba
-\frac{2 \gamma \phi}{r^4} + \frac{2 \gamma\phi'}{r^3} -\frac{\gamma \phi''}{r^2} -\frac{(2+3 \gamma)\phi'''}{r} -\frac{(1+\gamma)\phi''''}{2}=0 \ , \label{lineq1}
\ea
\ba
-\frac{(8+22\gamma) \phi}{r^4} - \frac{2 \gamma\phi'}{r^3} + \frac{(4+13 \gamma) \phi''}{r^2} -\frac{(2+5 \gamma)\phi'''}{r} -\frac{(1+3\gamma)\phi''''}{2}=0 \label{lineq2}
\ea
and 
\ba
-\frac{(4+10\gamma) \phi}{r^4} - \frac{2\gamma\phi'}{r^3} + \frac{(2+7 \gamma) \phi''}{r^2} -\frac{\gamma \phi'''}{r} -\frac{\gamma \phi''''}{2}=0 \label{lineq3}
\ea
respectively. The solution of Eq. \ref{lineq1} is given by
\ba
\phi= C_1 r^{\frac{-2\gamma}{1+\gamma}} + \frac{C_2}{r} + C_3 r + C_4 r^2  \label{sol} \ .
\ea
Plugging this solution, either in Eq. \ref{lineq2} or \ref{lineq3}, we get the same constraint over the constants which is given as
follows
\ba
(1+3 \gamma)\bigg[-(6\gamma^3+\gamma^2-5\gamma-2)C_1 + 2(1+\gamma)^4 r^{1+\frac{2 \gamma}{1+\gamma}} C_3\bigg] = 0 \label{constraint} \ .
\ea
Now, we have different solution for allowed values of $\gamma$ and other constants. The constraint Eq. \ref{constraint} gives one of 
the case where $\gamma=-1/3$. For this value we get
\ba
\phi= \frac{C_2}{r} + (C_1+C_3)r + C_4 r^2 \ ,
\ea
which is same solution as  in Eq. (\ref{conf_sol}) for the conformal theory.  However, for
$\gamma\ne -1/3$, we have $C_3=0$ and
\ba
6\gamma^3+\gamma^2-5\gamma-2=0 \ ,
\ea
which implies
\ba
\gamma= -\frac{2}{3} , \ \ \gamma =-\frac{1}{2} \ \ \mbox{and} \ \gamma=1 \ .
\ea
For these values of $\gamma$ the solutions are given by
\ba
\phi&=& C_1 r^4 + \frac{C_2}{r} + C_4 r^2 ,  \label{sol1} \\ 
    &=& \frac{C_2}{r} + (C_1+C_4) r^2 , \label{sol2} \\ 
    &=&  \frac{(C_1+ C_2)}{r} + C_4 r^2   \label{sol3}
\ea
respectively. The solution \ref{sol2} or \ref{sol3} with the data of galactic rotational curve \cite{Xue,Honma}
  for the Milky Way galaxy is plotted in Fig \ref{fig:vl}. For the large scale,  data is taken
 from the simulation II given in the table (3) of the Ref. \cite{Xue} and for small scale the data is taken from the table (2) 
of the  Ref. \cite{Honma}.
% \begin{table}[ht]
% \caption{Table for galactic rotational curve.}
% \label{milky_way}
% \centering
%     \begin{tabular}{|c|c|c|}
%         \hline
%         Distance   & Velocity v  & Error  \\
%         in Kpc &   (100 Km/sec) &  (100 Km/sec) \\
%  \hline
%         2.97 &2.03 & .225 \\ 
%         3.82 & 2.16 & .180 \\
%          4.67&2.26  & .185 \\
%         5.52 &2.18  &.090  \\
%          6.37 &2.27  &.040  \\
%           7.22 &2.25  &.045  \\
%           7.50&2.15  &.200  \\
%          8.07& 2.21 &.040  \\
%          12.5& 2.26 & .200 \\
%          17.5 &1.80  & .200 \\
%          22.5 & 1.64 & .200 \\
%          27.5 &1.83  & .200 \\
%           32.5&1.43  & .220 \\
%           37.5&1.83  & .390 \\
%          42.5 &2.03  & .350 \\
%           47.5 &1.66  & .300\\
%            55.0 &1.80  & .350\\
%         \hline
%     \end{tabular}
% \end{table}

 The effective velocity of star may be written as 
\ba
v^{2}= \frac{f}{2}\left(r\frac{\partial \phi}{\partial r}\right) \ ,
\ea
where $f = 9\times 10^{6}$ to make velocity unit as $(100 Km/sec)$.  For the best fit, the values of constants are 
\ba
C_2 = -7.39 \times 10^{-6} \  Kpc \ \ \mbox{and} \ \ (C_1+C_4)= 1.67 \times 10^{-10} \  Kpc^{-2} \ \ \mbox{for solution} \ (\ref{sol2}), \nonumber \\
(C_1+C_2) = -7.39 \times 10^{-6} \  Kpc \ \ \mbox{and} \ \ C_4=  1.67 \times 10^{-10} \  Kpc^{-2} \ \ \mbox{for solution} \  (\ref{sol3}).  
\ea 
The plot is shown by the dotted line. The further plot of the solution (\ref{sol1}) with solid line is shown in Fig. \ref{fig:vl}. The values of constants for this case are 
as following
\ba
C_1= - 5.16\times 10^{-14} \ {Kpc^{-4}}, \ C_2 = -7.22 \times 10^{-6} \ Kpc \ \ \mbox{and} \ \ C_4=3.78 \times 10^{-10} \ Kpc^{-2} \ .
\ea
The plot for the conformal theory is also shown with dashed line, where the gravitational potential is given by the 
Eq. (\ref{conf_sol}) and  for the best fit the values of the constants are given by
\ba
C_1= 2.6526 \times 10^{-6} \ Kpc, \ \ C_2=5.0460 \times 10^{-8} \ Kpc^{-1} \ \mbox{and} \ \ C_3=4.1366 \times 10^{-10} \ Kpc^{-2} \ .
\ea
The values of $\chi^{2}_{min}$
% \footnote{
% \ba
% \chi^2= \Sigma_{i} \left(\frac{theo_i-expt_i}{error}\right)^2\ea}
per degree of freedom for the best fit for the solution (\ref{conf_sol}), (\ref{sol1}) and (\ref{sol2})  are given by $3.19$, $5.54$ and $6.15$
respectively. However, the solution (\ref{sol1}) gives the best fit for the scale $> 15 Kpc$ as shown in the Fig. (\ref{fig:vl}). For
the large scale $> 15 Kpc$, we find  $\chi^{2}_{min}$ for the conformal theory; Eq. (\ref{conf_sol}) as $2.25$ whereas
for the case of unimodular gravity; Eq. (\ref{sol2}) and (\ref{sol1}) it is as
$1.09$ and $0.77$ respectively. Hence for the large scale, the theory of unimodular gravity describes the galactic rotational curve
with the best fit.

\section{Cosmological Solution} \label{Cosmo_sol}
It is known to us that Gauss-Bonnet action explains acceleration in the expansion of the universe \cite{Nojiri_Odin_Sasaki,
Zhou_Cope_Saffin,Ito_Nojiri} . Further, in the modified 
theory of gravity $f(R)=R^2$, we have exact de-Sitter solution \cite{Soti_Faraoni} for the vacuum. In this section, we test it explicitly 
as now the action has the unimodular
constraint in addition. For the FRW metric $[-1,a^2,a^2,a^2]$, where $a$ is scale factor of the universe, Eq. (\ref{field_eq2}) gives
the same equation for $0-0$ and $i-j$ components and it is given by
\ba
-(6+18\gamma)\left(\frac{a'}{a}\right)^4 + (9+27\gamma)\frac{a'^2 a''}{a^3}-(3+9\gamma)\left(\frac{a''}{a}\right)^2+
(1+3\gamma)\frac{a' a'''}{a^2}-(1+3\gamma)\frac{a''''}{a}=0 \label{eqn_scale1}\ .
\ea
 The Eq. (\ref{eqn_scale1}) may be written as  independent of the parameter $\gamma$ as
\ba
-6\left(\frac{a'}{a}\right)^4 + 9\frac{a'^2 a''}{a^3}-3\left(\frac{a''}{a}\right)^2+
\frac{a' a'''}{a^2}-\frac{a''''}{a}=0 \label{eqn_scale2} \ .
\ea
Looking  over Eq. (\ref{eqn_scale2}), one may conclude for the exact de-Sitter solution,
\ba
a=a_0 e^{H_0 t}\ ,
\ea
which explain the acceleration in the expansion of universe, where $H_0$ is Hubble constant. Hence, the de Sitter solution satisfies
both the conformal theory and the theory of unimodular gravity.
\section{Discussion and Conclusions} \label{conclusion}
A scale invariant model of higher order invariant in the action is presented. 
The unimodular constraint on the  theory is also considered. Scale invariance allows only quadratic terms of curvature
scalar in the action, whereas consideration of unimodular theory in addition constrain on the values of the parameter of 
the resulting theory. It is shown that for the parameter $\gamma=-1/2$ and $1$,  the solution of the gravitational potential
includes the Schwarzschild solution as well as the term corresponding to the integration constant. The solution for this 
case explains the galactic rotational curve, but the corresponding gravitational field increases as distance increases whereas for $\gamma=-2/3$,
 the solution has one more term proportional to $r^4$ so that the velocity or corresponding gravitational field decreases after $\sim 42 \ Kpc$. Furthermore, 
  the solution of conformal theory is recovered for $\gamma=-1/3$. The conformal solution has a lighter bump at $\sim 30 \ Kpc$. Hence, the unimodular 
theory of gravity has good behavior for the large scale rather than that of conformal theory. The proper scale
 invariant matter source term in the action might describe the rotational curve for the 
low range. We will proceed it further in the future publication. The theory is interesting as it does not 
require the dark matter which has not been observed in the solar neighborhood so far. Furthermore, the de Sitter solution is also obtained 
for the considered theory explaining the dynamics of current era.

\bigskip
\noindent
{\bf Acknowledgements} \\

I thank  Subhendra Mohanty,  Pankaj Jain and Girish Chakrabarty for useful discussions.

\begin{spacing}{1}
\begin{small}

\end{small}
\end{spacing}
\end{document}